\documentclass[a4paper,12pt]{article}
\pdfoutput=1
\usepackage{graphicx, rotating}
\usepackage{hyperref}
\usepackage{slashed}
\usepackage{ifpdf}
\usepackage{amssymb}
\usepackage{multicol}

\ifx\pdfoutput\undefined
\usepackage[dvips,bookmarks=false]{hyperref}	
\else
\usepackage{hyperref}	
\fi
\def\hhref#1{\href{http://arxiv.org/abs/#1}{#1}} 

\usepackage{amsfonts}
\usepackage{amsmath}
\usepackage{slashed}
\usepackage{soul}
\usepackage{multirow}

\newcommand{\beq}{\begin{equation}}
\newcommand{\eeq}{\end{equation}}

\newcount\Mac  \Mac=1  
\newcommand{\ifMac}[2]{\ifnum\Mac=1 #1 \else #2 \fi}
\def\putps(#1,#2)(#3,#4)#5#6{\ifnum\Mac=1 \put(#1,#2){\special{picture #5}}
\else  \put(#3,#4){\includegraphics{#6}} \fi}

\newcommand{\One}{\hbox{1\kern-.24em I}}

\newcommand{\lascia}[1]{}
\makeatletter

\hypersetup{
colorlinks=true,
linkcolor=blue,
anchorcolor=black,
citecolor=green,
filecolor=cyan,
menucolor=rossos,
runcolor=filecolor,
urlcolor=rossos,
bookmarks=true,
bookmarksnumbered=true
}

\def\art{\@ifnextchar[{\eart}{\oart}}
\def\eart[#1]#2#3#4#5#6{{\rm #2}, {#3 #4} {\rm (#6) #5} [arXiv:{\hhref{#1}}]}

\def\hepart[#1]#2{{\rm #2, [arXiv:\hhref{#1}]}}
\newcommand{\oart}[5]{{\rm #1}, {#2 #3} {\rm (#5) #4}}

\numberwithin{equation}{section}

\newcounter{alphaequation}[equation]
\def\thealphaequation{\theequation\hbox to
0.6em{\hfil\alph{alphaequation}\hfil}}
\def\eqnsystem#1{
\def\@eqnnum{{\rm (\thealphaequation)}}
\def\@@eqncr{\let\@tempa\relax \ifcase\@eqcnt \def\@tempa{& & &} \or
  \def\@tempa{& &}\or \def\@tempa{&}\fi\@tempa
  \if@eqnsw\@eqnnum\refstepcounter{alphaequation}\fi
\global\@eqnswtrue\global\@eqcnt=0\cr}
\refstepcounter{equation} \let\@currentlabel\theequation \def\@tempb{#1}
\ifx\@tempb\empty\else\label{#1}\fi
\refstepcounter{alphaequation}
\let\@currentlabel\thealphaequation
\global\@eqnswtrue\global\@eqcnt=0 \tabskip\@centering\let\\=\@eqncr
$$\halign to \displaywidth\bgroup \@eqnsel\hskip\@centering
$\displaystyle\tabskip\z@{##}$&\global\@eqcnt\@ne
\hskip2\arraycolsep\hfil${##}$\hfil& \global\@eqcnt\tw@\hskip2\arraycolsep
$\displaystyle\tabskip\z@{##}$\hfil
\tabskip\@centering&\llap{##}\tabskip\z@\cr}

\def\endeqnsystem{\@@eqncr\egroup$$\global\@ignoretrue} \makeatother

\def\Tr{\mathop{\rm Tr}}
\def\circa#1{\,\raise.3ex\hbox{$#1$\kern-.75em\lower1ex\hbox{$\sim$}}\,}

\usepackage{multicol}
\usepackage{color}
\definecolor{rosso}{cmyk}{0,1,1,0.4}
\definecolor{rossos}{cmyk}{0,1,1,0.55}
\definecolor{rossoc}{cmyk}{0,1,1,0.2}
\definecolor{blu}{cmyk}{1,1,0,0.3}
\definecolor{blus}{cmyk}{1,1,0,0.6}
\definecolor{bluc}{cmyk}{1,1,0,0.1}
\definecolor{verde}{cmyk}{0.92,0,0.59,0.25}
\definecolor{verdec}{cmyk}{0.92,0,0.59,0.15}
\definecolor{verdes}{cmyk}{0.92,0,0.59,0.4}
\definecolor{grigio}{cmyk}{0,0,0,0.07}
\definecolor{rosa}{cmyk}{0,0.1,0.1,0.02}
\definecolor{rosino}{cmyk}{0,0.05,0.05,0.02}
\definecolor{rosas}{cmyk}{0,0.3,0.25,0.05}
\definecolor{celeste}{cmyk}{0.1,0,0,0.02}
\definecolor{giallino}{cmyk}{0,0,0.4,0.02}
\definecolor{rosso}{cmyk}{0,1,1,0.4}
\definecolor{rossos}{cmyk}{0,1,1,0.55}
\definecolor{rossoc}{cmyk}{0,1,1,0.2}
\definecolor{blu}{cmyk}{1,1,0,0.3}
\definecolor{bluc}{cmyk}{1,1,0,0.1}
\definecolor{blucc}{cmyk}{0.7,0.5,0,0}
\definecolor{viola}{cmyk}{0,1,0,0.6}
\definecolor{viola2}{cmyk}{0,1,0.2,0.6}
\definecolor{verde}{cmyk}{0.92,0,0.59,0.25}
\definecolor{verdec}{cmyk}{0.92,0,0.59,0.15}
\definecolor{verdes}{cmyk}{0.92,0,0.59,0.4}
\definecolor{verdino}{cmyk}{0.12,0,0.09,0.05}
\definecolor{giallo}{cmyk}{0,0,1,0}
\definecolor{gialloverde}{cmyk}{0.44,0,0.74,0}

\newfam\rsfsfam
\def\mathscr#1{{\fam\rsfsfam\relax#1}}

\newcommand{\MET}{E\llap{/\kern1.5pt}_T}

\begin{document}
\hfill IFT-UAM/CSIC-11-80\\
\color{black}
\vspace{0.5cm}
\begin{center}
{\Huge\bf Strong, weak and flavor scalar triplets\vspace{3mm}\\ for the CDF $Wjj$ anomaly}\\
\bigskip\color{black}\vspace{0.6cm}
{{\large\bf Duccio Pappadopulo$^{a}$, Riccardo Torre$^{b}$}
} \\[7mm]
{\it  (a) Institut de Th\'eorie des Ph\'enom\`enes Physiques, EPFL,  CH--1015 Lausanne, Switzerland}\\[3mm]
{\it  (b) Instituto de F\'isica Te\'orica UAM/CSIC, Calle Nicol\'as Cabrera 13-15, E-28049 Madrid, Spain}\\[3mm]
\end{center}
\bigskip
\centerline{\large\bf\color{blus} Abstract}
\begin{quote}
A model describing the $4.1\sigma$ $Wjj$ anomaly observed by the CDF experiment at the Tevatron collider is introduced. It features new scalar particles which are charged both under the $SU(3)_{C}$ and the $SU(2)_{L}$ gauge groups and which couple to pairs of quarks. We introduce several identical replicas of the scalar multiplets in order to leave an unbroken $U(3)_Q\times U(3)_U\times U(3)_D$ flavor symmetry to satisfy the constraints coming from flavor physics. We discuss the LHC reach on the new scalar resonances both in the resonant production channel (with the $Wjj$ final state) and in the QCD pair production channel (with the $4j$ final state). 
\normalsize
\color{black}
\end{quote}


\section{Introduction}
The CDF collaboration presented the analysis of the invariant mass distribution of jet pairs produced in association with a (leptonically decaying) $W$ boson in a sample corresponding to a total integrated luminosity of $4.3$ fb$^{-1}$ \cite{CDFCollaboration:2011vq}\footnote{Further details of this analysis can be found in Ref.~\cite{Cavaliere:um}}. The study displays an excess of events in the $120-160$ GeV range. The significance of the excess, which at the moment does not find any conventional explanation within the Standard Model (SM), was determined to be at the level of 3.2 standard deviations\footnote{This number is obtained fitting the signal to a gaussian distribution.}. A recent increase in the integrated luminosity \cite{Annovi:uy} used in the analysis has strengthened the significance of the anomaly to 4.1$\sigma$. A similar analysis carried out by the D\O~collaboration does not confirm the CDF result \cite{DCollaboration:2011wm}. Waiting for the LHC to confirm or disprove this anomaly, it is interesting to explore the consistency and plausibility of explanations involving physics Beyond the SM (BSM). Various attempts to explain the anomaly in terms of BSM physics are already present in the literature: $Z'$  models \cite{2011arXiv1105.2699E,Chen:2011wp,Wang:2011wt,Carpenter:2011yj}, models with new scalar fields coupled to quarks \cite{Yu:2011cw,Buckley:2011ww,Cheung:2011un,Wang:2011wh,Anchordoqui:2011wg}, two Higgs doublet models \cite{Gunion:2011bx} and other quirky explanations \cite{Harnik:2011mv}.

Our attempt fits in the di-quarks category. We are going to introduce colored scalar fields $\Phi$ charged under $SU(2)_{L}$, coupled to quark pairs schematically as
\begin{equation}\label{interaction}
g_{Q}\,\Phi\cdot {\rm quark} \cdot {\rm quark}.
\end{equation}

In the absence of Electro-Weak Symmetry Breaking (EWSB) the scalars are resonantly produced and they promptly decay into pairs of jets. 
Decay channels containing $W$ bosons, necessary to explain the anomaly, appear if a mass splitting within an $SU(2)_{L}$ mutiplet or mixing among states with different $SU(2)_{L}$ quantum numbers are present after EWSB.

This article is organized as follows. In Section \ref{sec:ourmodel} we introduce our model. We motivate the choice of the matter content and coupling structure discussing, in a qualitative way, the bounds coming from Electro-Weak (EW) and flavor physics. In Section \ref{sec:constraints} we enter a detailed discussion of the bounds on the couplings and masses of the new particles. We compare the parameter space allowed by existing physics with the one where the CDF excess is explained.
In Section \ref{sec:lhcpheno} we present the prospects of LHC to re-discover the signal and we briefly discuss other promising signatures. We conclude in Section \ref{sec:conclusions}.

\section{The model}\label{sec:ourmodel}
\begin{figure}[!t]
\begin{center}
\includegraphics[scale=0.52]{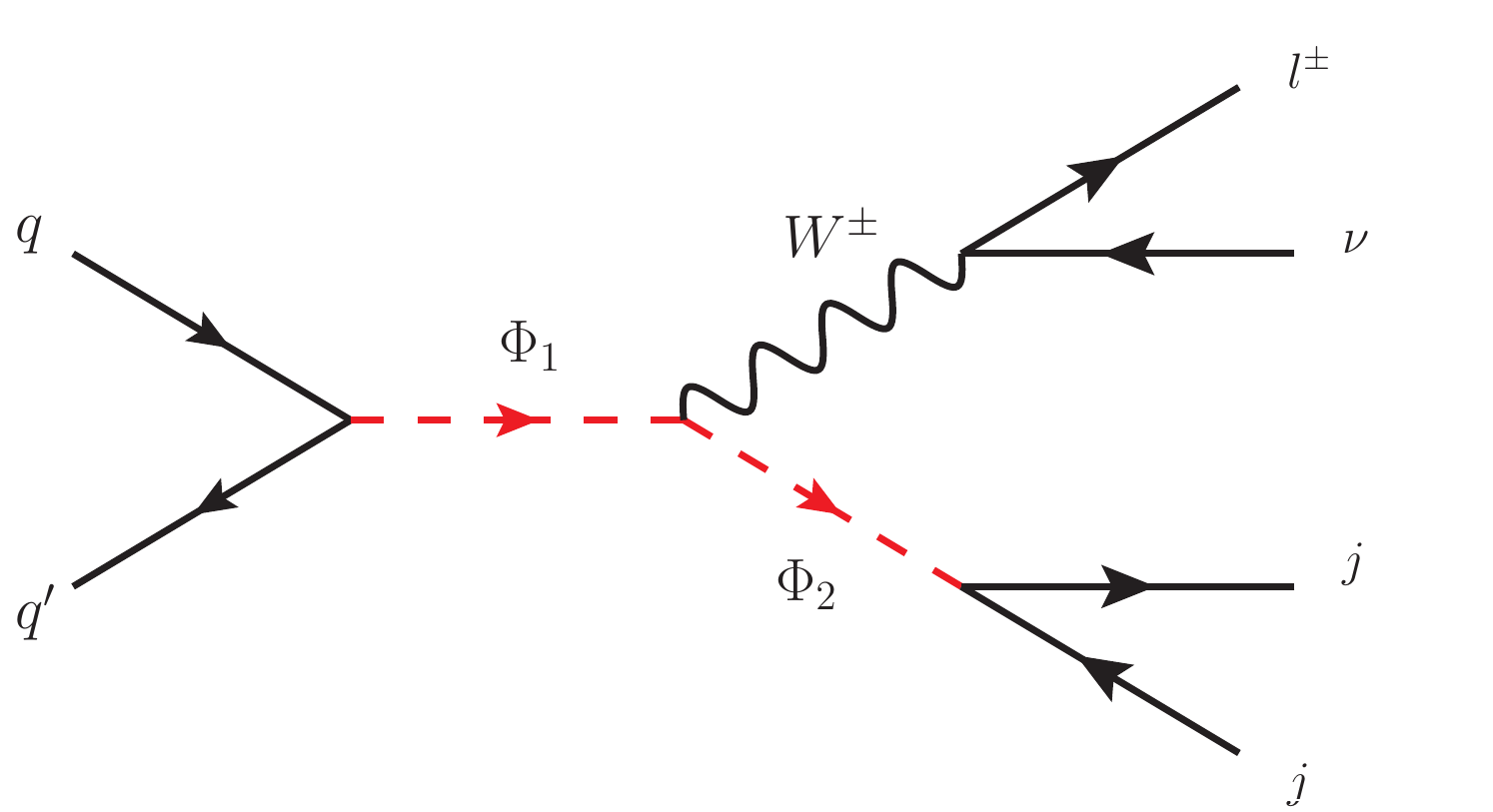}   
\end{center}
\caption{\label{resonance} 
\small
Feynman diagram representation of the resonant production through quark-quark annihilation of a scalar particle $\Phi_{1}$ decaying into a (leptonically decaying) $W^{\pm}$ boson and a (hadronically decaying) scalar particle $\Phi_{2}$.
}
\end{figure}

Our strategy to reproduce the CDF signal is explained in Fig.~\ref{resonance}. 
We will introduce a scalar multiplet resonantly produced through its coupling to quarks and decaying into a lighter state through a charged current interaction. The focus on resonant production is to allow for a smaller Yukawa coupling to quarks.
%

The set of possible di-quarks resonances is readily obtained. We start listing the usual SM particles quantum numbers in Tab.~\ref{smfermions}. We then look at the irreducible representations of the gauge and global SM symmetry groups appearing in the Lorentz invariant product of two quark fields. Such reduction is shown in Tab.~\ref{diquarks}. We restrict the table to three out of the six possible combinations since these are the only one which contain non-singlet $SU(2)_{L}$ representations.
Without loss of generality we take the new scalar multiplets to belong to the conjugate representations of those appearing in Tab.~\ref{diquarks}.

\begin{table}[t]
\begin{center}
\begin{tabular}{ccccccc}
&$U(3)_Q$ &$U(3)_U$&$U(3)_D$&$SU(3)_C$&$SU(2)_L$&$U(1)_Y$\\ \hline
$Q$&$\mathbf 3$&$\mathbf 1$&$\mathbf 1$&$\mathbf 3$&$\mathbf 2$&1/6\\
$U$&$\mathbf 1$&$\mathbf {\bar 3}$&$\mathbf 1$&$\mathbf {\bar 3}$&$\mathbf 1$&-2/3\\
$D$&$\mathbf 1$&$\mathbf 1$&$\mathbf{\bar 3}$&$\mathbf {\bar 3}$&$\mathbf 1$&1/3\\
\hline
\end{tabular}
\end{center}
\caption{\small\label{smfermions}The SM fermion content, together with its quantum number assignments. All spinors are left-handed Weyl spinors. We also show the $U(3)_Q\times U(3)_U\times U(3)_D$ flavor quantum numbers.}
\end{table}
\begin{table}[t]
\begin{center}
\begin{tabular}{ccccccc}
&$U(3)_Q$ &$U(3)_U$&$U(3)_D$&$SU(3)_C$&$SU(2)_L$&$U(1)_Y$\\ \hline
$QQ$&$\mathbf {\bar 3}_A \oplus \mathbf 6_S$&$\mathbf 1$&$\mathbf 1$&$\mathbf {\bar 3}_A \oplus \mathbf 6_S$&$ \mathbf 3_S$&1/3\\
$QU$&$\mathbf 3$&$\mathbf {\bar 3}$&$\mathbf 1$&$\mathbf  1\oplus \mathbf 8$&$\mathbf 2$&-1/2\\
$QD$&$\mathbf 3$&$\mathbf 1$&$\mathbf{\bar 3}$&$\mathbf  1\oplus \mathbf 8$&$\mathbf 2$&1/2\\
\hline
\end{tabular}
\end{center}
\caption{\small\label{diquarks}List of di-quarks with charged current interactions.~The $S/A$ index specifies the symmetry/antisymmetry of the representation.}
\end{table}

Without making any assumption on the flavor structure of the model, the Yukawa interactions in Eq.~\eqref{interaction} would be strongly constrained by the bounds on flavor violating observables. A recent discussion about these constraints in model with di-quarks can be found in Ref.~\cite{Giudice:2011ak}. To easily avoid such bounds we demand the new scalars to respect some flavor symmetry that we can use to implement Minimal Flavor Violation (MFV) \cite{DAmbrosio:2002kc}. The simplest choice is to enforce the full $U(3)_Q\times U(3)_U\times U(3)_D$ SM flavor symmetry. If we require the di-quarks to fill complete representations of this flavor group we are left with the 6 possibilities shown in Tab.~\ref{diquarks}. Notice that the $QQ$ di-quarks  allows for two possibilities, one of which is totally symmetric in EW, color and flavor indices and the other one that is antisymmetric in flavor and color indices. We choose the smallest one, i.e. the antisymmetric.

The last crucial ingredient in the model is the introduction of a splitting among the different electric charge components of the scalar multiplets. The simplest way to achieve this is to introduce a direct coupling with the Higgs doublet \cite{Vecchi:2011un}
\begin{eqnarray}\label{quartichiggs}
\lambda_{H^2} \, (\Phi^\dagger T^A_\Phi\Phi)\, (H^\dagger \tau^A H)\,,
\end{eqnarray}
where $\tau^A=\sigma^A/2$ and $T_\Phi^A$ are the $SU(2)_{L}$ generators in the representation of $\Phi$. 
The coupling in Eq.~\eqref{quartichiggs} splits the components of $\Phi$ according to ($v\approx$ 246 GeV)
\begin{equation}\label{splitting1}
\Delta M^2_\Phi=\lambda_{H^2}\frac{v^2}{2} T_\Phi^3.
\end{equation}

The mass splitting in Eq.~\eqref{splitting1} breaks custodial symmetry and generates a correction to the $\hat T$ parameter at loop level. In the limit of small relative splitting the size of the contribution can be estimated noticing that $\lambda_{H^2}$ represents a $\Delta I=1$ violation of custodial isospin. The leading correction to $\hat T$ is thus of the order
\begin{equation}\label{estimateT}
\hat T\approx \mathcal O(1)N_F N_C\frac{g^2}{16\pi^2}\frac{(\Delta M^2_\Phi)^{2}}{M_W^2 M_\Phi^2}.
\end{equation}
The mass splitting between the states in the multiplet has to be larger than $M_W$ in order to reproduce the signal through resonant production of $\Phi$. Furthermore the lowest lying state in the multiplet must have a mass around 150 GeV to reproduce the shape of the di-jet peak observed by the CDF collaboration. With these inputs the right-hand side of Eq.~\eqref{estimateT} is too big to be acceptable. A careful evaluation of $\hat T$, which is by the way needed due to the large relative mass splitting inside the multiplet, confirms this (see Appendix \ref{Tparam}).

An alternative way to reproduce the signal while avoiding the above problems with $\hat T$ is to introduce two states $\Phi_1$ and $\Phi_2$, the first heavier than the second, which belongs to different $SU(2)_{L}$ representations before EWSB and are mixed after. The mixing is introduced by couplings to the Higgs doublet which, at the renormalizable level, can be either cubic or quartic. The flavor and color quantum numbers of the new state are fixed to be the same of the old one.

The possible $SU(2)_L\times U(1)_Y$ representations can be catalogued in a simple way. For cubic couplings the weak-isospin of the $\Phi_1$ state has to be equal to $\left|I_2\pm\frac{1}{2}\right|$ (if $I_2=0$ then $I_1=\frac{1}{2}$) while its hypercharge can be either $Y_2+\frac{1}{2}$ or $Y_2-\frac{1}{2}$ due to the pseudo-real nature of the $SU(2)$ representations.

In case quartic couplings are used, the $\Phi_1^\dagger\Phi_2$ combination must couple to one of the three possible triplet combinations of two Higgs doublets (in order for any EWSB mass splitting to be generated). This means that the possible $SU(2)_L\times U(1)_Y$ assignments for $\Phi_1$ are among the states with weak-isospin $\left| I_2-1\right|$, $I_2$, $I_2+1$ (if $I_2=0$ then $I_1=1$) with hypercharge $Y_2$, $Y_2\pm 1$.

This discussion immediately implies that cubic and quartic couplings mixing $\Phi_1$ with $\Phi_2$ cannot coexist. The spectrum of possible quantum numbers and couplings is still fairly wide. Some more or less motivated assumptions can be introduced. One is to require that no $Zjj$ excess is generated at a comparable rate to the $Wjj$ one. This requires the $Z$ boson to couple diagonally to the new states after mixing\footnote{This assumption will imply the BR $\Phi_1\to Z\Phi_2$ to be vanishing, but cannot forbid the $Z\Phi$ associated production. This will be however suppressed with respect to the resonant production.}. The necessary and sufficient condition for this to happen is to have mixing among states with the same hypercharge. This excludes cubic couplings with the Higgs doublet and forces the quartic couplings to be with the real triplet $H^\dagger\tau^A H$ combination. 

Still various possibilities remain. The color adjoint representations may be dropped due to their larger size. The $QU$ color singlet di-quarks have already been used in Ref.~\cite{Nelson:2011up} to give a joint explanation of CDF and $t\bar t$ asymmetry anomaly, though using a smaller flavor group.

A compelling and minimal possibility which emerges from Tab.~\ref{diquarks} and from the previous discussion is thus represented by a $QQ$ di-quark with hypercharge $-1/3$ which is a triplet under the flavor, color and isospin groups. This state is assumed to mix with an isospin singlet with the same hypercharge, flavor and color. Notice that with this minimal choice for the state to mix with we automatically get rid of extra coupling of the electroweak singlet to the quarks.

\subsection{The Lagrangian}\label{subsec:thelagrangian}
We introduce two states\footnote{The notation for the representations is $(SU(3)_F,SU(3)_C,SU(2)_L)_Y$.}: $T=(\mathbf{3},\mathbf{3},\mathbf 3)_{-1/3}$ and $S=(\mathbf{ 3},\mathbf{ 3},\mathbf 1)_{-1/3}$. To distinguish the $SU(2)_{L}$ components of $T$ we adopt a matrix notation
\begin{equation}
T=\begin{pmatrix}T_{-1/3}/\sqrt{2}&T_{2/3}\\ T_{-4/3}&-T_{-1/3}/\sqrt{2}\end{pmatrix}.
\end{equation}
All the flavor and color indices are suppressed and the lower indices on $T$ are the electric charges of the different components of the $SU(2)_{L}$ triplet. $T$ transforms under $SU(2)_L$ rotations as $T\to L T L^\dagger$.

Using the Weyl notation for spinors, the Lagrangian for $T$ and $S$ is written as follows
\begin{eqnarray}
\notag\mathcal L&=& \Tr[D_\mu T^\dagger D^\mu T]-M_T^2\Tr[T^\dagger T]+\Tr[D_\mu S^\dagger D^\mu S]-M_S^2\Tr[S^\dagger S]\\
&-&\lambda_{1}\, H^\dagger T T^\dagger H-\lambda_{2}\, H^\dagger T^\dagger T H-\lambda_3\, S^\dagger S H^\dagger H-\lambda\, H^\dagger T H S^\dagger +{\rm h.c.}\\
\notag&+&g_Q \epsilon_{ijk}\epsilon_{\alpha\beta\gamma}(\sigma_2 Q^{i\alpha})^T T^{j\beta} Q^{k\gamma}+{\rm h.c.}\,,\label{diquarkcouplings}
\end{eqnarray}
where the trace is over the weak-isospin indices\footnote{When the flavor and color contractions are not completely obvious we indicate flavor indices with latin letters (e.g. $i,l,\ldots$) and color ones with greek letters (e.g. $\alpha,\beta,\ldots$).}.
All the other cubic and quartic couplings we have omitted in the Lagrangian~\eqref{diquarkcouplings} are not relevant to us as long as they lead to a stable ground state where color is not spontaneously broken.
The mass spectrum is determined after EWSB by the mass matrix
\begin{equation*}
\Delta \mathcal L_{\rm mass}=- V^\dagger \mathcal M V,\quad V^T=(T_{2/3},T_{-4/3}, T_{-1/3}, S)\,,\\
\end{equation*}
\begin{equation}\label{massmatrix}
\mathcal M=\begin{pmatrix}
M_T^2+\lambda_2 v^2/2 & 0&0 &0\\
0& M_T^2+\lambda_1 v^2/2 & 0 &0 \\
0 & 0 &M_T^2+ (\lambda_1+\lambda_2) v^2/4 &\lambda v^2/2\sqrt 2\\
0 &0&\lambda v^2/2\sqrt 2& M_S^2 +\lambda_3 v^2/2
\end{pmatrix}\,.
\end{equation}

Unless otherwise stated, we assume in the following that the $\lambda_{1,2,3}$ couplings are negligible\footnote{$\lambda_1$ and $\lambda_2$ contribute to the $\hat T$ parameter as explained in Section \ref{sec:ourmodel} and must be small.}. With these specifications the model has 4 independent parameters, two masses ($M_T$ and $M_S$) and 2 couplings ($\lambda$ and $g_Q$). The spectrum is determined by $\lambda$ which mixes $T_{-1/3}$ with $S$. $T_{2/3}$ and $T_{-4/3}$ are mass eigenstates with mass $M_T$ while 
\begin{equation}\label{rottriplet}
\begin{pmatrix}S_1\\S_2\end{pmatrix}=\begin{pmatrix}
\cos\theta_T&\sin\theta_T\\-\sin\theta_T&\cos\theta_T\end{pmatrix}\begin{pmatrix}T_{-1/3}\\S\end{pmatrix},\qquad \tan2\theta_T=\frac{2 M_{TS}^2}{M_T^2-M_S^2},
\end{equation}
where $M_{TS}^2=\lambda v^2/2\sqrt 2$ and 
\begin{equation}
M_{S_{1,2}}^2=\frac{M_T^2+M_S^2\pm \sqrt{(M_T^2-M_S^2)^2+4 M_{TS}^2}}{2}.
\end{equation}
The mass $M_{S}$, which is unphysical, can be traded for the mass of the lightest state, which we assume to be $S_{2}$, that is the di-jet invariant mass around which the CDF excess is observed. In the gauge basis we can expand the interaction with the quarks in order to show explicitly the coupling of every $SU(2)_{L}$ component
\begin{equation}\label{Lintgaugebasis}
g_Q \epsilon_{ijk}\epsilon_{\alpha\beta\gamma}(\sigma_2 Q^{i\alpha})^T T^{j\beta} Q^{k\gamma}=g_Q \epsilon_{ijk}\epsilon_{\alpha\beta\gamma}(-i \sqrt{2}Q_D^{i\alpha}T_{-1/3}^{j\beta}Q_U^{k\gamma}-i Q_D^{i\alpha}T_{2/3}^{j\beta}Q_D^{k\gamma}+i Q_U^{i\alpha}T_{-4/3}^{j\beta}Q_U^{k\gamma}).
\end{equation}
When we rotate to the quarks mass basis the Cabibbo-Kobayashi-Maskawa (CKM) matrix enters in Eq.~\ref{Lintgaugebasis} through the  substitution
\begin{equation}
Q_U^i\to Q_U^j V_{CKM}^{*ji}.
\end{equation}

\section{Constraints}\label{sec:constraints}
\begin{figure}[!t]
\begin{center}
\includegraphics[scale=1.1]{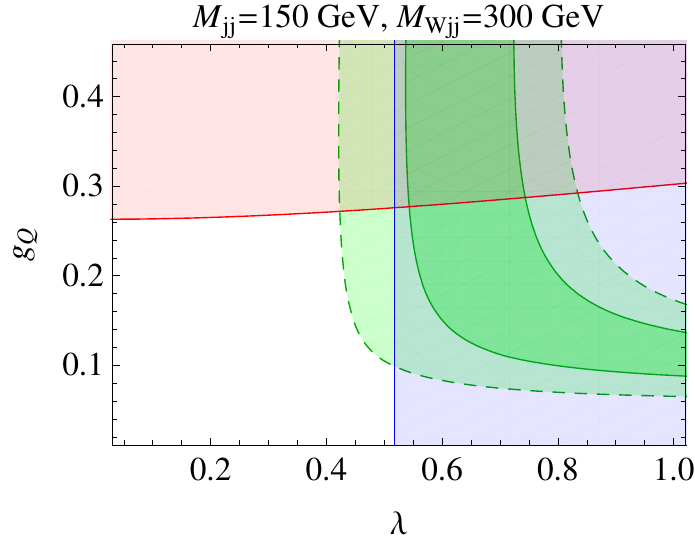}
\end{center}
\vspace{-4mm}\caption{\small Collection of bounds from CDF di-jet search (red, horizontal line) and 1-loop contribution to $\Delta \hat T$ (blue, vertical line) together with the region favored by the $Wjj$ CDF excess (green, 1$\sigma$ full contour, 2$\sigma$ dashed contour), in the $(\lambda,\, g_Q)$ plane. The experimental efficiencies and acceptances for the $Wjj$ analysis are included as explained in the text. }\label{bounds} 
\end{figure}

\begin{figure}[!t]
\begin{center}
\includegraphics[scale=1.2]{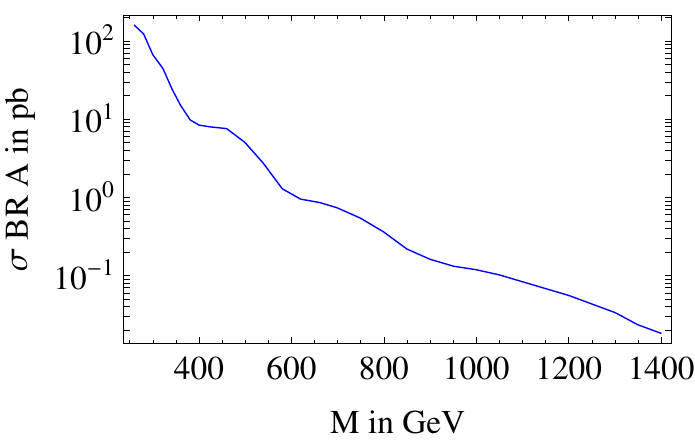}   
\end{center}
\vspace{-6mm}\caption{\small The di-jet CDF bound, as in Fig.~1a in Ref.~\cite{CDFCollaboration:2009hy}.}\label{rippedbound} 
\end{figure}

There are two major constraints on the model. The first comes from the corrections to the $\hat T$ parameter which are induced by the mixing of the triplet with the singlet. We give their detailed expression in Appendix \ref{Tparam}. A second relevant constraint comes from the direct search performed by the CDF collaboration in the di-jet final state \cite{CDFCollaboration:2009hy}. The strongest limits are from the resonant production of the heavier scalar states\footnote{CDF searches are not effective for masses below $260$ GeV. The old UA2 data are less constraining and we don't include them.} which then decay into a pair of jets. We show these two constraints in Fig.~\ref{bounds}. To draw the di-jet Tevatron exclusion we use the bound from Fig.~1a in Ref.~\cite{CDFCollaboration:2009hy} (which is calculated for resonant production and decay of a $W'$ vector boson) and we assume, conservatively, all the acceptances to be equal to 1. The Tevatron exclusion is reported in Fig.~\ref{rippedbound}. For the bound on $\hat T$ we use $-3\cdot 10^{-3}\leq\hat T\leq 3\cdot 10^{-3}$ \cite{afterlep2} neglecting the correlation with $\hat S$, since the latter is vanishing in our model.

We include in Fig.~\ref{bounds} the favored region in the $(\lambda,\,g_Q)$ parameter space which explains the CDF excess for $M_{jj}=150$ GeV and $M_T=300$ GeV. This region is obtained reproducing the total number of observed leptons in $7.3$ fb$^{-1}$ of data\footnote{In the CDF analysis this number is the result of a gaussian fit to the excess.} \cite{Annovi:uy} (240$\pm$55 electrons and 158$\pm$45 muons), applying the cuts in Tab.~\ref{kincuts}. 
\begin{table}[t]
\begin{center}
\begin{tabular}{c|c}
{\bf Variable} & {\bf Cut}\\ \hline
$p_{T}^{l}$	&  $\geq 20$ GeV\\
$\MET$		&  $\geq 25$ GeV\\
$p_{T}^{j}$	&  $\geq 20$ GeV\\
$p_{T}^{jj}$	& $\geq 40$ GeV\\
$|\eta_{j}|$	& $\leq 2.4$\\
$|\Delta \eta_{jj}|$	& $\leq 2.5$\\
$|\Delta \phi_{\MET,j_{1}}|$& $\geq 0.4$\\
\hline
\end{tabular}
\end{center}
\caption{\small\label{kincuts} Kinematic cuts used to reproduce the CDF $Wjj$ signal.}
\end{table}
We estimate the acceptances corresponding to these cuts through a parton level analysis obtained implementing our model in {\sc MadGraph5} \cite{Alwall:2011uj}. We show the results for the acceptances in Tab.~\ref{signalaccTEV} for different values of the relevant masses. The importance of experimental efficiencies, i.e.~reconstruction efficiencies, cannot be overstated. To get a na\"ive idea of these efficiencies we follow the simple method discussed in Appendix \ref{CDFefficiencies}. Using the expected diboson events in the electron and muon samples from the analysis of Ref.~\cite{CDFCollaboration:2009hy} we extract
\begin{equation}
\epsilon_e=0.14,\qquad  \epsilon_\mu=0.12.
\end{equation}
The final number we use in Fig.~\ref{bounds} is thus
\begin{equation}
N_\ell=\sum_{i=e,\,\mu} [\sigma \cdot {\rm BR}] _{\ell_i\nu_ijj}\,\mathcal A\, \epsilon_i
\end{equation}
where $\mathcal A$ and $\epsilon$ are the geometrical acceptance and experimental efficiencies.

\begin{table}[t]
\begin{center}
\begin{tabular}{rc|cccccc}
 & & \multicolumn{6}{c}{\small $m_{S_2}$ (GeV)} \\
 & &$265$ &$300$&$350$&$400$&$450$&$500$\\ \hline
\multirow{3}{*}{\small $m_{S_2}$ (GeV)}&130& $0.52$&$0.61$&$0.67$&$0.71$&$0.75$&$0.75$\\
&150& $0.44$&$0.57$&$0.67$&$0.72$&$0.76$&$0.78$\\
&170& $0.22$&$0.52$&$0.65$&$0.72$&$0.76$&$0.79$
\end{tabular}
\end{center}
\caption{\small\label{signalaccTEV} Signal acceptances at the Tevatron ($\sqrt{s}=1.96$ {\rm TeV}) for the kinematic requirements of Table \ref{kincuts} for different masses (in {\rm GeV}) of the heavy and light scalars.}
\end{table}

The $p\bar p\to W jj$ cross section needed to explain the signal is roughly $4$ pb. As the plot shows a tension with $\hat T$ remains. If we allow $\hat T$ to be as big as $4\cdot 10^{-3}$, which could be motivated by an additional negative contribution coming from a heavy Higgs boson, then the 1$\sigma$ region explaining the CDF excess becomes allowed. Since the contribution to $\hat T$ turns out to be positive the agreement of the model with the Electro-Weak Precision Tests (EWPT) could be improved with a heavier Higgs boson\footnote{An additional possibility that could be explored to ameliorate the tension with $\hat T$ is to mix the new scalar states through a coupling which violates custodial isospin by a smaller amount, $\Delta I =1/2$, so that in the small $\Delta M^2/M^2$ limit $\Delta \hat T$ would be suppressed by a higher power of $\Delta M^2/M^2$. This is realized for instance if the triplet mixes with a doublet through a cubic coupling with the Higgs boson. Since the hypercharge of the mixed states cannot be the same this possibility generically introduces a non vanishing $\Phi_1\to \Phi_2 Z$ branching ratio.}. Notice that due to resonant production, the couplings needed to explain the anomaly are small and perturbative. This implies, in particular, that the Yukawa coupling $g_Q$ stays perturbative up to very large scales despite the large matter content of the model\footnote{The running of $g_Q$ is determined by $dg_Q/d\log \mu=g_Q^3/2\pi^2$. Only wave-function renormalization contribute to the running of $g_Q$.}. We find that for values of $M_T$ slightly above $350$ GeV we are unable to satisfy the EW constraints whilst explaining the CDF signal.

\subsection{Flavor physics}\label{subsec:flavor}
The only sources of flavor violation in the model are the usual Yukawa matrices and the model respects the MFV hypothesis. This automatically guarantees the absence of tree-level Flavor Changing Neutral Currents (FCNC). This is shown switching off the interactions of the quarks with the $T_{-1/3}$ components of the triplets and those with the $W$ boson. The Lagrangian in Eq.~\eqref{Lintgaugebasis} is then invariant under independent unitary rotations of both $Q_U$ and $Q_D$ if one rotates at the same time also $T_{2/3}$ and $T_{-4/3}$.

\begin{figure}[!t]
\begin{center}
\hspace{2mm}\includegraphics[scale=0.72]{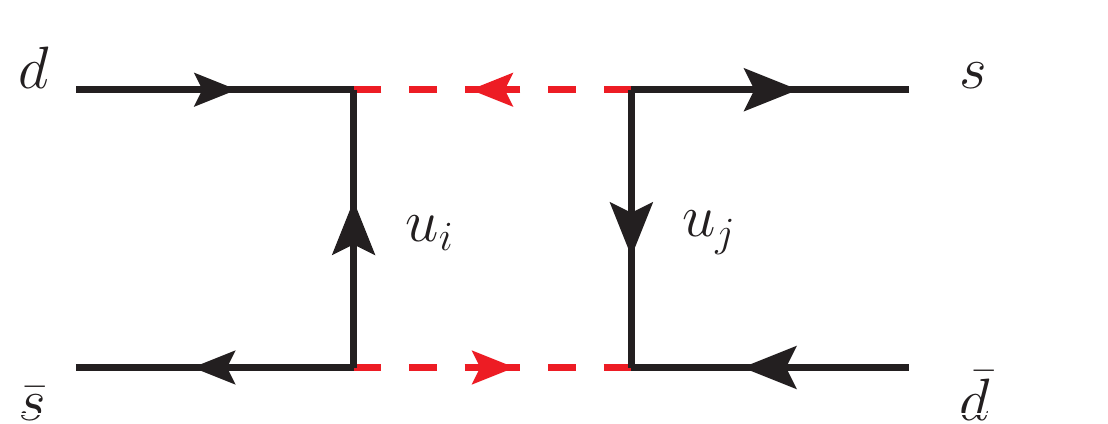}\hspace{5mm}\includegraphics[scale=0.72]{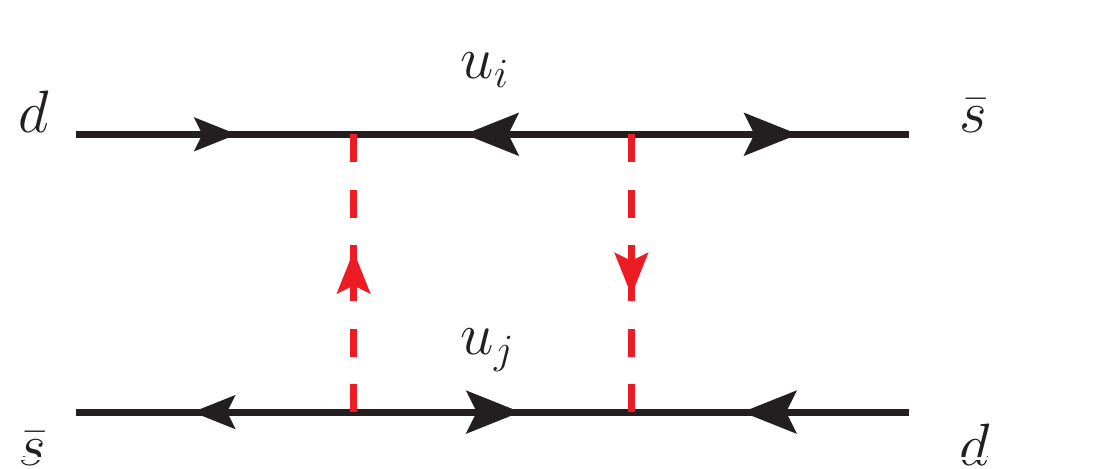}
\end{center}
\caption{\label{1loopFCNC} 
\small
Box diagrams contributing at 1-loop to the $K$-$\overline K$ mixing.
}
\end{figure}

Such suppression of tree-level FCNC appears generically if the di-quark couples antisymmetrically in flavor space \cite{Giudice:2011ak, Vecchi:2011un}. This happens since at the 2 generation level, its couplings are proportional to the $SU(2)$ invariant tensor $\epsilon_{ij}$ and they respect an accidental $SU(2)$ flavor symmetry. This accidental symmetry is however not enough to suppress loop-induced FCNC as the one depicted in Fig.~\ref{1loopFCNC}, since the third generation can propagate in the loop\footnote{Single scalar exchange boxes vanish.}.

Enforcing MFV through the Lagrangian in Eq.~\eqref{Lintgaugebasis}, provides a GIM mechanism which guarantees the diagram in Fig.~\ref{1loopFCNC} to be null for vanishing quark masses. Considering for instance the contribution to the $K$-$\overline K$ mixing Hamiltonian, one generates the operator
\begin{equation}
C_K^1 \,\bar s^\alpha\gamma^\mu d^\alpha\,\bar s^\beta\gamma_\mu d^\beta
\end{equation}
with a coefficient which is given by (at leading order in $m_t^2/M_T^2$ and neglecting the mixing with the singlet)
\begin{equation}
C_K^1\approx\frac{g_Q^4 (V_{31}^* V_{32})^2}{8\pi^2}\frac{m_t^2}{M_T^4}=\left(\frac{1}{1.5\cdot 10^5{\,\rm TeV} }\right)^2\left(\frac{g_Q}{0.3}\right)^4\left(\frac{300{\,\rm GeV}}{M_T}\right)^4.
\end{equation}
This contribution is completely safe if compared with the current experimental bounds (see Tab.~IV of Ref.~\cite{UTfitCollaboration:2007p3234}). Notice in particular that no extra sources of CP violation arise from $g_Q$ or $\lambda$, since complex phases in these couplings can always be reabsorbed through a field redefinition of $T$ and $S$. Similar considerations apply to $B_d$-$\overline B_d$ and $B_s$-$\overline B_s$ transitions. 

Other flavor constraints arise considering $\Delta B=1$ transitions, either at tree-level or at 1-loop order. Starting with the former, due to the flavor symmetry, the only interesting effects come from the exchange of the charge $-1/3$ mass eigenstates. This are incorporated in an effective Hamiltonian which in the quark mass basis looks like\footnote{We used the Fierz identity \cite{Dreiner:2008tw} 
\begin{equation}
2(\chi_1^\dagger\chi_3^\dagger)(\chi_4\chi_2)=\bar \Psi_1\gamma^\mu\Psi_2\, \bar\Psi_3\gamma_4\Psi_4
\end{equation}
being $\chi_i$ and $\Psi_i$ a Weyl and Dirac spinors respectively, related by $\chi_i=\frac{1-\gamma_5}{2}\Psi_i$.
}
\begin{equation}\label{deltab1}
g_Q^2V^{*j3}V^{hi}\left(\frac{\cos^2\theta_T}{M_1^2}+\frac{\sin^2\theta_T}{M_2^2}\right)(\bar b^\gamma\gamma^\mu d^{i\alpha}\, \bar u^{h\alpha}\gamma_\mu u^{j\gamma}-\bar b^\gamma\gamma^\mu d^{i\gamma}\, \bar u^{h\alpha}\gamma_\mu u^{j\alpha}).
\end{equation}
The simplest requirement is to ask the total $b$ quark width not to exceed the measured one. Taking $\tau_b=(1.64\pm 0.01)\cdot 10^{-12}$ s as a rough average of the various $B$ meson lifetimes, one gets that the relative New Physics (NP) contribution to the $b$ width $\Delta \Gamma_b/\Gamma$ must not exceed $\sim1\%$. Computing $\Delta\Gamma_b$ using Eq.~\eqref{deltab1} one gets
\begin{equation}
\Delta\Gamma_b\approx\frac{g_Q^4 |V_{23}|^2 m_b^5}{256\pi^3}\left(\frac{\cos^2\theta_T}{M_1^2}+\frac{\sin^2\theta_T}{M_2^2}\right)^2,
\end{equation}
to be compared with the naive $b$ quark width to get
\begin{equation}
\frac{\Delta \Gamma_b}{\Gamma_b}\approx \frac{g_Q^4}{6}\left(\frac{v^2\cos^2\theta_T}{M_1^2}+\frac{v^2\sin^2\theta_T}{M_2^2}\right)^2\lesssim 1\%.
\end{equation}
This never constitutes a problem for the parameter range of interest.

Other less trivial checks come from $|\Delta B|=|\Delta S|=1$ transitions such as $B\to \phi_s K$ or $B\to J/\psi K$. We will follow the recent analysis of similar constraints contained in Ref.~\cite{Vecchi:2011un}.  We assume the NP contributions to the Wilson coefficients $C_{J/\psi}(m_b)$ and $C_{\phi_s}(m_b)$ defined in Eq.~16,17 of Ref.~\cite{Lunghi:2001af} not to exceed 10\% of their SM value. The boundary value at the scale $M_1\approx M_2\approx M_W$ of these coefficients is read from Eq.~\eqref{deltab1}. Assuming for simplicity $\theta_T=0$ we get
\begin{eqnarray}
\frac{\delta C_{J/\psi}}{C_{J/\psi}}:&&\left(\frac{g_Q}{0.3}\right)^2\left(\frac{300\, {\rm GeV}}{M_T}\right)^2\lesssim 1.7\\
\frac{\delta C_{\phi_s}}{C_{\phi_s}}:&& \left(\frac{g_Q}{0.3}\right)^2\left(\frac{300\, {\rm GeV}}{M_T}\right)^2\lesssim 1.3,
\end{eqnarray}
which are not particularly severe constraints on the parameter space of the model.

As a last check one can calculate the contributions to the $b\to s\gamma$ branching ratio, which is induced by penguin diagrams with $T_{-1/3}$ circulating in the loop. Here we report the expression for the Wilson coefficient $\delta C_7(M_W)$ defined by the effective Hamiltonian
\begin{equation}
\mathcal H= -\frac{4 G_F}{\sqrt{2}} V_{32}^*V_{33} (C^{SM}_7+\delta C_7)\, \frac{em_b}{16\pi^2} \bar s^\alpha\sigma_{\mu\nu} F^{\mu\nu} P_R b^\alpha.
\end{equation}
The experimental bound is $\delta C_7(m_W)\lesssim 0.1$ \cite{Hurth:2008p3281}. We neglect the running from the mass of the scalar to $m_W$ and we assume no mixing between the singlet and the triplet. 
We obtain, for $m_t/M_T\lesssim 1$,
\begin{equation}
\delta C_7\approx 0.01 \left(\frac{g_Q}{0.3}\right)^2\left(\frac{300\, {\rm GeV}}{M_T}\right)^4,
\end{equation}¥
which is completely safe.

\subsection{Modification of the Higgs boson couplings}\label{subsec:higgs}

The scalar multiplets introduced to explain the CDF excess have the peculiar couplings to the Higgs boson described  by the interactions in Eq.~\eqref{diquarkcouplings}. At the one-loop level these interactions will introduce a coupling of the SM Higgs boson to gluons and photons similarly to what happens for the top quark. In the $m_h\ll M_{T,S}$ limit the leading effect is a contribution to the $h G_{\mu\nu}^aG^{a\mu\nu}$, $h F_{\mu\nu}F^{\mu\nu}$  operators. The former is relevant for the Higgs boson production cross section, while the second influences its decay width into photons.

The calculation of these effects goes as follows \cite{Low:2009gl}. Consider a matter multiplet of mass $M$, made of scalars or fermions, such that $m_h\ll M$. The leading dependence of the low energy effective action on the Higgs field comes, in this limit, from the Higgs dependence of the $M$ threshold in the running of the QCD (QED) gauge coupling. In terms of its UV value, the coupling constant at a low scale $\mu$ is given by
\begin{equation}\label{gaugehiggs}
\mathcal{L}\supset-\frac{1}{4}\left(\frac{1}{g_S(\mu_{UV})^2}+\frac{b_2}{16\pi^2}\log\frac{\mu_{UV}^2}{M(h)^2}+\frac{b_1}{16\pi^2}\log\frac{M(h)^2}{\mu^2}\right)G_{\mu\nu}^aG^{a\mu\nu}
\end{equation}
where $M(h)^2$ has to be understood as the mass matrix in the $h$ background with the logarithm defined accordingly. The structure of the $hGG$ coupling in the canonical basis follows from Eq.~\eqref{gaugehiggs}
\begin{equation}
\mathcal{L}\supset\frac{g_S^2}{64\pi^2}(b_2-b_1)\left[\frac{\partial}{\partial h} \log M(h)^2\right]_{h=0} hG_{\mu\nu}^aG^{a\mu\nu}.
\end{equation}
A complex scalar (Dirac fermion) in a representation $R$ contributes to $b_2-b_1$ as $1/3\,(4/3) C(R)$ where $C(R)$ is the Dynkin index of the representation.

Using the mass matrix in Eq.~\eqref{massmatrix} the calculation is straightforward. We drop for simplicity all the quartic couplings except the one proportional to $\lambda$ (which is the one relevant to explain the CDF anomaly). We express the result as the ratio $\delta \mathcal A_{gg(\gamma\gamma)}/\mathcal A_{gg(\gamma\gamma)}^{SM}$ where $\mathcal A_{gg(\gamma\gamma)}$ is the coefficient of the operator $(h/v) GG\,(FF)$ either in the SM on in the model we are proposing
\begin{eqnarray}
\frac{\delta \mathcal A_{gg}}{\mathcal A_{gg}^{SM}}&=&-\frac{3}{8}\frac{\lambda^2 v^4}{M_S^2M_T^2-\lambda^2 v^4/8} ,\\
\frac{\delta \mathcal A_{\gamma\gamma}}{\mathcal A_{\gamma\gamma}^{SM}}&=&-\frac{3}{64}\frac{\lambda^2 v^4}{M_S^2M_T^2-\lambda^2 v^4/8}.
\end{eqnarray}
Taking $\lambda=0.6$ and fixing $M_T$ and $M_S$ to reproduce Fig.~\ref{bounds} we get $\delta \mathcal A_{gg}/\mathcal A_{gg}^{SM}\approx 0.25$ which amounts in a reduction of the $gg\to h$ cross section of roughly 40\%. The modification of the $h\to \gamma\gamma$ amplitude is negligible.

\section{LHC phenomenology}\label{sec:lhcpheno}
The new scalar particles we introduced to explain the CDF $Wjj$ anomaly are expected to give a signal at the LHC. 
We find the most interesting topologies to be the single production of the heavy scalar decaying to a $Wjj$ final state and the QCD pair production of the scalars followed by decays into $4j$, $W+4j$ and $2W+4j$ final states.

\begin{figure}[!t]
\begin{center}
\includegraphics[scale=1.1]{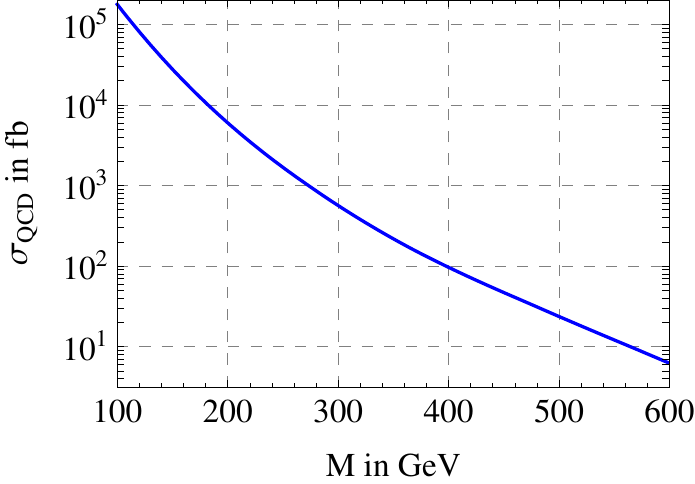}~~~~~\includegraphics[scale=1.1]{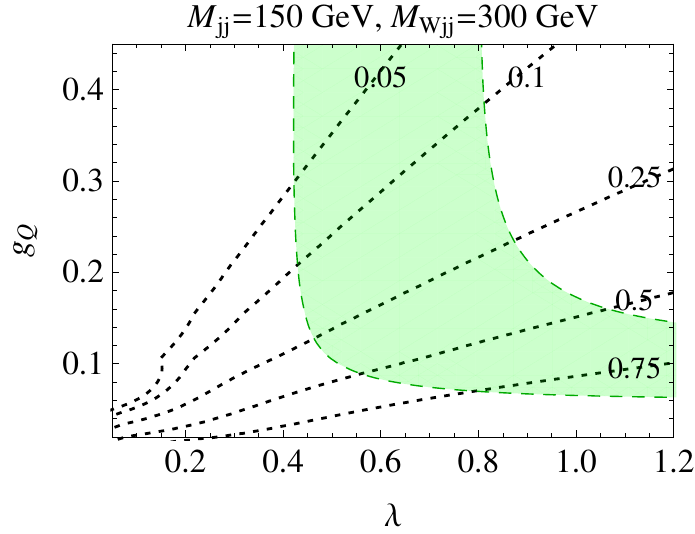} 
\end{center}
\vspace{-4mm}\caption{\small Left panel: leading-order QCD pair production cross section for a color triplet scalar of a given mass at the LHC with $\sqrt{s}=7$ TeV. Right panel: dashed contours corresponding to the average branching ratio into $W$ bosons for all the $T_{-4/3}$, $T_{2/3}$ particles in the model. The green region is the one reproducing the CDF excess at the 2$\sigma$ level for the parameters shown in the picture.}\label{figBRsigma} 
\end{figure}

We will analyze the resonant production in the next paragraph. For the QCD pair production rate, some order of magnitude estimate of the signal cross section can be obtained from the two panels of Fig.~\ref{figBRsigma}.

Recently the ATLAS Collaboration has performed a search for colored scalars in the $4j$ final state with an integrated luminosity of $34$ pb$^{-1}$  \cite{Aad:2011yh}. The bound applies to scalars with a mass ranging from 100 to 200 GeV. Assuming equal signal acceptances we can use the ATLAS result to estimate the experimental sensitivity to the pair production of the lightest scalar states in the model.

From Fig.~3 of Ref.~\cite{Aad:2011yh} the $95\%$ CL upper bound on the QCD pair production cross section times the BR is $\sim 3\cdot 10^{5}$ fb for a $150$ GeV scalar. This experimental value has to be compared with the total production cross section of three color triplet scalar of 150 GeV mass decaying only to jets. This number is obtained from Fig.~\ref{figBRsigma}, and is roughly $9\cdot 10^{4}$ fb. Assuming the experimental sensitivity to scale like $1/\sqrt{L}$, a ten times bigger luminosity, roughly 0.4 fb$^{-1}$, is required by ATLAS to exclude the model at the 95\% CL. Such integrated luminosity is already available to the collaboration.

For what concerns the heavy scalars, from Fig.~\ref{figBRsigma} we see that a single color triplet of $300$ GeV mass is QCD pair produced with a cross section around $550$ fb. To get a $4j$ signal we can sum the cross section of all the 9 heavier scalars which we assume to be degenerate. Adopting an average 75\% BR into di-jets, the total $4j$ signal is slightly above 2.5\,pb. The signal for di-quark pair productions into a $4j$ final state has been studied in Ref.~\cite{DelNobile:1199750} where it was shown that a signal of a $400$ GeV di-quark (with unit BR into jets) can emerge from the SM background even with $1$ fb$^{-1}$ of integrated luminosity (see Fig.~4 of Ref.~\cite{DelNobile:1199750}). We expect the signal of a lighter scalar di-quark, but with a smaller BR into jets, to emerge with a comparable statistics. Slightly softer cuts should be applied on the jet $p_{T}$ in order to optimize the signal acceptance in this case, due to the smaller scalar mass. 

If $W$ bosons are required in the signal, only 6 states are allowed to be pair produced (namely the 3 flavor of the charge $-4/3$ and $2/3$ scalars). Assuming again a degenerate 300\,GeV scalar multiplet and an average 75\% BR into jets, the $W+4j$ and the $2W+4j$ signals are respectively around 1.3\,pb and 0.2\,pb. 
Without $b$-tagging top quark production associated with jets is a large irreducible background to both signals. We find in particular that $t\bar t$ and $t\bar t + jj$ amounts respectively to 90\,pb and 48\,pb at the LHC with $\sqrt{s}=7$ TeV\footnote{These backgrounds are calculated with {\sc{MadGraph5}}. For the $t\bar t + jj$ background default cut for jet identification are used ($p^{\min}_{Tj}=20$\,GeV, $\eta_j^{\max}=5$, $\Delta R^{\min}_{jj}=0.4$).}. Given the discouraging appearance of these numbers further studies regarding the relevance of $b$-tagging or the existence of taylored cuts to reduce the relative importance of the background are required to say whether these channels are of any relevance for the next year of the LHC.

\subsection{$Wjj$ final state at the LHC}\label{subsec:lhcwjj}
\begin{figure}[!t]
\begin{center}
\includegraphics[scale=1.1]{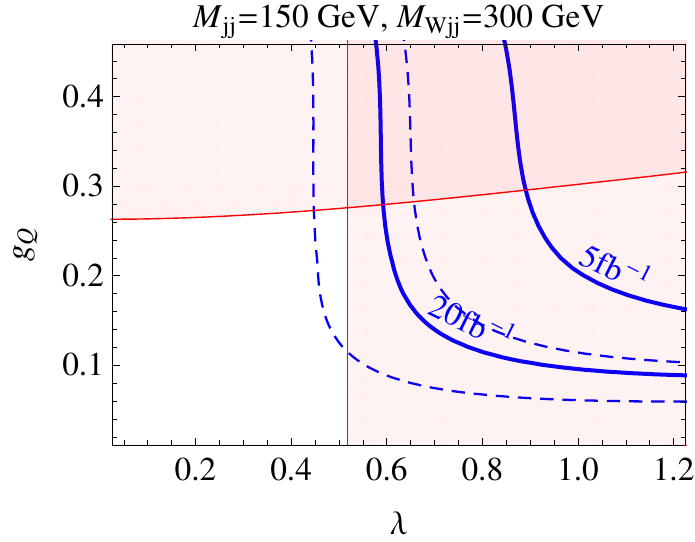}
\end{center}
\vspace{-4mm}\caption{\small 
The blue contours in the $(\lambda, g_Q)$ plane indicate where the minimal cross section in the $\ell\nu jj$ channel needed for a 5\,$\sigma$ (full) or $3\sigma$ (dashed) discovery is achieved for a given integrated luminosity ($5, 20\,{\rm fb}^{-1}$) at the LHC with $\sqrt{s}=7$ TeV. The red bands are the Tevatron dijet and $\hat T$ constraints as in Fig.~\ref{bounds}.
}\label{LHC7} 
\end{figure}

As a direct test of the model as a possible explanation of the CDF anomaly, the existence of the new scalar particles can be observed at the LHC in the same channel used by the CDF experiment at the Tevatron, looking for an excess of events in the di-jet invariant mass region between $120$ and $160$ GeV. We perform a pure statistical analysis, without any attempt to take into account systematic uncertainties or detector effects. Our analysis is done at parton level. In order to give an idea of the feasibility of this search, we discuss not only the significance of the signal, but also the signal over background ratio, which is of fundamental importance when the background has large systematic uncertainties as in the case of the QCD $Wjj$ background. We write the signal significance and the signal over background ratio as
\begin{equation}\label{significanceLHC}
\begin{array}{l}
\displaystyle s=\frac{\sigma_{\text{signal}}[pp\to W(\ell\nu)S_{2}( jj)]\times L \times \mathcal{A}_{\text{signal}} \times \epsilon_{\text{signal}}}{\sqrt{\sigma_{\text{SM}}[pp\to \ell\nu jj]\times L \times \mathcal{A}_{\text{SM}} \times \epsilon_{\text{SM}}}}\,,\\\\
\displaystyle r=\frac{\sigma_{\text{signal}}[pp\to W(\ell\nu)S_{2}( jj)]\times L \times \mathcal{A}_{\text{signal}} \times \epsilon_{\text{signal}}}{\sigma_{\text{SM}}[pp\to \ell\nu jj]\times L \times \mathcal{A}_{\text{SM}} \times \epsilon_{\text{SM}}}\,,
\end{array}
\end{equation} 
where $\sigma_{i}$, $\mathcal{A}_{i}$ and $\epsilon_{i}$, $i=$ signal,SM are the total cross sections, the kinematic acceptances and the detector efficiencies for the $\ell\nu jj$ final state, coming respectively from the signal and from the SM background. $L$ is the integrated luminosity. The kinematic acceptance corresponds to the cuts of Table \ref{kincuts} plus a cut on the invariant mass of the two jets $120<M_{jj}<160$ GeV. Assuming the efficiencies of the signal and the background to be the same we can simplify the expressions \eqref{significanceLHC} to
\begin{equation}\label{significanceLHC2}
\begin{array}{l}
\displaystyle s=\frac{\sigma_{\text{signal}}[pp\to W(\ell\nu)S_{2}( jj)]\times \mathcal{A}_{\text{signal}}}{\sqrt{\sigma_{\text{SM}}[pp\to \ell\nu jj]\times \mathcal{A}_{\text{SM}}}}
\sqrt{L\times \epsilon}\,,\\\\
\displaystyle r=\frac{\sigma_{\text{signal}}[pp\to W(\ell\nu)S_{2}( jj)]\times \mathcal{A}_{\text{signal}}}{\sigma_{\text{SM}}[pp\to \ell\nu jj] \times \mathcal{A}_{\text{SM}}}\,.
\end{array}
\end{equation} 
The SM background for our kinematic requirements has been computed using {\sc MadGraph5} and is
\begin{equation}\label{backgr}
\sigma_{\text{SM}}[pp\to \ell\nu jj]\times \mathcal{A}_{\text{SM}}=33\,\text{pb}\,.
\end{equation}
We take as the reference kinematic acceptance for the signal the one for $M_{T}=300$ GeV and $M_{S_{1}}=150$ GeV, which is, for the LHC with $\sqrt{s}=7$ TeV, $ \mathcal{A}_{\text{signal}}=0.47$. This acceptance is slightly below the corresponding value obtained at the Tevatron for the same masses (see Table \ref{signalaccTEV}) due to the different c.o.m.~energy. We expect, at the LHC, the kinematic acceptance to change for different values of the masses of the mother and the daughter resonances in a similar way to that described by Table \ref{signalaccTEV} for the Tevatron. To be conservative we assume a relatively small value for the detector efficiencies at the LHC (of the same order of the CDF one), namely $\epsilon = 0.1$ both for the electrons and the muons. We can therefore define the minimum signal cross section which is expected to give rise to a $5\sigma$ ($3\sigma$) discovery with a given integrated luminosity $L$ as
\begin{equation}\label{minsignal}
\sigma_{\text{signal}}^{\text{min}}[pp\to W S_2]\approx 30\,(18)\,{\rm pb}\,\sqrt{\frac{1\,{\rm fb^{-1}}}{L}}\,.
\end{equation}
with and expected minimum signal over background ratio
\begin{equation}\label{sigoverbackmin}
r^{\text{min}}
\approx 0.1\,(0.06)\,\sqrt{\frac{1\,{\rm fb^{-1}}}{L}}\,,
\end{equation}
where we substituted $BR(W\to l\nu)\approx 0.22$ for $l=e,\mu$ and $BR(S_{2}\to jj)\approx 1$.
The total cross section \eqref{minsignal}, which is the minimum cross section required to have a signal significance of $5\,(3)$ standard deviations, can be converted in allowed regions in the $(\lambda,g_{Q})$ plane as a function of the integrated luminosity $L$ as shown in Fig.~\ref{LHC7}. This superficial (and optimistic) analysis shows that a 5$\sigma$ signal significance can be attained at the end of 2012 (where the predicted total integrated luminosity is around 20\,${\rm fb}^{-1}$) only allowing some extra negative contribution to the $\hat T$ parameter (e.g. from an heavy Higgs).


\section{Conclusions}\label{sec:conclusions}
We discussed a model which can explain the excess of events observed by the CDF experiment in the $Wjj$ channel. The model introduces 3 families of scalar fields, each family made of 2 color triplet $SU(2)_{L}$ representations: a singlet and a triplet. The EW triplets have Yukawa couplings to the left-handed quark fields. Their $T_3=0$ components are split from the  others through a mixing with the singlets. The CDF signal is reproduced through the resonant production of the heavier un-mixed states which then decay into the lightest, mostly singlet, component through the emission of a W boson. Flavor bounds are avoided implementing a global $U(3)^3$ flavor symmetry explicitly broken only by the SM Yukawa interactions. In the region of the parameter space where the CDF excess is reproduced we find a mild tension with the $\hat T$ parameter. This could in principle be softened by some extra negative contribution to $\hat T$ coming for instance from a heavy Higgs boson with a somewhat reduced production cross section.

The re-discovery of the signal within the first run of the LHC (7\,\text{TeV}, 20\,\text{fb}$^{-1}$, end 2012) may be challenging. The most promising signatures are the same $Wjj$ signal observed by CDF along with the $4j$ one resulting from QCD pair production of the new scalars followed by their hadronic decay. The latter become the most relevant one in case of a light Higgs boson and no extra negative contribution to $\hat T$.

\section*{Acknowledgments}
We thank Riccardo Rattazzi for the original idea leading to this work. We thank Andrea de Simone for collaborating at the time this project was started and Roberto Franceschini for discussing with us. R.T. thanks Paolo Francavilla for discussion about the LHC phenomenology. The work of D.P. is supported by the Swiss National Science Foundation under contracts No.~200021-116372 and the work of R.T has been partially supported by the Swiss National Science Foundation under contracts No.~200021-12523, by the European Commission under the contract PITN-GA-2010-264564 LHCPhenoNet, by the Spanish MICINN under grants CPAN CSD2007-00042 (Consolider-Ingenio 2010 Programme) and  FPA2010-17747 and  by the Community of Madrid under the grant HEPHACOS S2009/ESP-1473. D.P. thanks the Kavli Institute of Theoretical Physics where part of this work was carried through.

\appendix

\section{$\hat{T}$ parameter from the splitting}\label{Tparam}
The contribution to the $\hat T$ parameter from a complex $SU(2)$ triplet whose isospin components are splitted according to equation \ref{splitting1} is given by
\begin{equation}
\frac{g^2 N_CM_T^2}{8\pi^2 M_W^2}\left(1-\frac{{\rm arctan}\, x}{x}- \log (1-x^2) \right),
\end{equation}
where $x\equiv\Delta M^2/M_T^2$ and $N_C$ is the number of colors. Assuming on the other hand the splitting among the components of the EW triplet to occur through the mixing of an isospin singlet with the $T_3=0$ component of the triplet (see Eq.~\eqref{diquarkcouplings}) the contribution to $\hat T$ is given by
\begin{equation}
\frac{g^2 N_CM_T^2}{8\pi^2 M_W^2}\left(1+\frac{x_1 \log x_1}{1-x_1} \cos^2\theta_T+\frac{x_2 \log x_2}{1-x_2} \sin^2\theta_T\right),
\end{equation}
where the notation is the same as in Eq.~\eqref{rottriplet} and $x_{1,2}\equiv M^2_{1,2}/M_T^2$.

\section{CDF efficiency}\label{CDFefficiencies}
We want to compare our model with the CDF excess without doing any collider simulation. Thus we have to extract from the CDF data their detector efficiencies for electron and muon identification. We make a distinction between geometrical/kinematical acceptances and detection efficiencies. The firsts are due to the cuts in Tab.~\ref{kincuts} while the seconds to detector effects. Several na\"ive approximations are made assuming that these two effects can be disentangled. For instance we assume the first to be independent of the nature of the lepton, and the second not to depend at all on the process under exam. 
To extract the efficiencies we use the data in Ref.~\cite{CDFCollaboration:2011vq} for the expected number of di-boson events \cite{annovi_private}. These numbers have been computed from the numbers of observed events and the ratios of the observed over expected events for the electron and muon samples given in Table 1 of Ref.~\cite{CDFCollaboration:2011vq}. We find 156 expected electrons and 97 expected muons in 4.3 fb$^{-1}$. Following our simplified assumptions we write these two numbers as
\begin{equation}\label{eqefficiencies}
N_{e/\mu}=\mathcal L \cdot(\sigma_{WW} \cdot\mathcal A_{WW\to e/\mu+X}+\sigma_{WZ}\cdot \mathcal A_{WZ\to e/\mu+X})\cdot\epsilon_{e/\mu}
\end{equation}
where $\sigma_{VV}$ are the di-boson cross section inclusive of the $K$-factor, i.e. $\sigma_{WW}=11.66$\,pb and  $\sigma_{WZ}=3.46$\,pb \cite{Campbell:387805}, $\mathcal A$ are the relative geometric acceptances\footnote{We include in these acceptances the branching ratio to a final state with a lepton of definite flavor.} and $\epsilon_{e/\mu}$ are the detector efficiencies we want to calculate. Once the acceptances are known, from Eq.~\eqref{eqefficiencies} we can extract $\epsilon_{e/\mu}$. We estimate these acceptances using a simple parton level Montecarlo obtaining
\begin{equation}
A_{WW\to e/\mu+X}=3.2\%,\qquad  A_{WZ\to e/\mu+X}=2.0\%.
\end{equation}
Substituting these acceptances, the cross sections, the luminosity and the expected number of electron and muons into Eq.~\eqref{eqefficiencies} we find the efficiencies
\begin{equation}
\epsilon_e=0.14,\qquad  \epsilon_\mu=0.12.
\end{equation}

\bibliography{draft}{}

\providecommand{\href}[2]{#2}\begingroup\raggedright\begin{thebibliography}{10}

\bibitem{CDFCollaboration:2011vq}
{\bf CDF} Collaboration, T.~Aaltonen {\em et.~al.}, {\it Invariant {M}ass
  {D}istribution of {J}et {P}airs {P}roduced in {A}ssociation with a ${W}$
  boson in $p\bar{p}$ {C}ollisions at $\sqrt{s}=1.96$ {T}e{V}},  {\em Phys.
  Rev. Lett.} {\bf 106} (2011) 171801,
  [\href{http://xxx.lanl.gov/abs/1104.0699}{{\tt arXiv:1104.0699}}] [\href{http://inspirebeta.net/record/894999}{Inspire}].

\bibitem{Cavaliere:um}
V.~Cavaliere, {\it Measurement of ${WW} + {WZ}$ {P}roduction {C}ross {S}ection
  and {S}tudy of the {D}ijet {M}ass {S}pectrum in the $l\nu$ + jets {F}inal
  {S}tate at {CDF}},  {\em FERMILAB-THESIS-2010-51} (2010) [\href{http://inspirebeta.net/record/881126}{Inspire}].

\bibitem{Annovi:uy}
A.~Annovi, P.~Catastini, V.~Cavaliere, and L.~Ristori, {\it Invariant {M}ass
  {D}istribution of {J}et {P}airs {P}roduced in {A}ssociation with a ${W}$
  boson in $p\bar{p}$ {C}ollisions at $\sqrt{s}=1.96$ {T}e{V}}, .
  [\href{http://www-cdf.fnal.gov/physics/ewk/2011/wjj/}{4.3 fb$^{-1}$}]
  [\href{http://www-cdf.fnal.gov/physics/ewk/2011/wjj/7_3.html}{7.3
  fb$^{-1}$}].

\bibitem{DCollaboration:2011wm}
{\bf D\O} Collaboration, V.~M. Abazov {\em et.~al.}, {\it Bounds on an
  anomalous dijet resonance in ${W}$+jets production in $p\bar{p}$ collisions
  at $\sqrt{s}=1.96$ {T}e{V}},  {\em Phys. Rev. Lett.} {\bf 107} (2011) 011804,
  [\href{http://xxx.lanl.gov/abs/1106.1921}{{\tt arXiv:1106.1921}}] [\href{http://inspirebeta.net/record/913398}{Inspire}].

\bibitem{2011arXiv1105.2699E}
T.~Enkhbat, X.-G. He, Y.~Mimura, and H.~Yokoya, {\it Colored {S}calars {A}nd
  {T}he {CDF} ${W}+$dijet {E}xcess},
  \href{http://xxx.lanl.gov/abs/1105.2699}{{\tt arXiv:1105.2699}} [\href{http://inspirebeta.net/record/899593}{Inspire}].

\bibitem{Chen:2011wp}
C.-H. Chen, C.-W. Chiang, T.~Nomura, and Y.~Fusheng, {\it A light charged
  {H}iggs boson in two-{H}iggs doublet model for {CDF} ${W}jj$ anomaly},
  \href{http://xxx.lanl.gov/abs/1105.2870}{{\tt arXiv:1105.2870}} [\href{http://inspirebeta.net/record/900204}{Inspire}].

\bibitem{Wang:2011wt}
X.-P. Wang, Y.-K. Wang, B.~Xiao, J.~Xu, and S.~hua Zhu, {\it New color-octet
  vector boson revisited},  {\em Phys. Rev.} {\bf D 83} (2011) 115010,
  [\href{http://xxx.lanl.gov/abs/1104.1917}{{\tt arXiv:1104.1917}}] [\href{http://inspirebeta.net/record/895687}{Inspire}].

\bibitem{Carpenter:2011yj}
L.~M. Carpenter and S.~Mantry, {\it {C}olor-{O}ctet-{E}lectroweak-{D}oublet
  {S}calars and the {CDF} {D}ijet {A}nomaly},  {\em Phys. Lett.} {\bf B 703}
  (2011) 479--485, [\href{http://xxx.lanl.gov/abs/1104.5528}{{\tt
  arXiv:1104.5528}}] [\href{http://inspirebeta.net/record/897689}{Inspire}].

\bibitem{Yu:2011cw}
F.~Yu, {\it A {Z}' {M}odel for the {CDF} {D}ijet {A}nomaly},  {\em Phys. Rev.}
  {\bf D 83} (2011) 094028, [\href{http://xxx.lanl.gov/abs/1104.0243}{{\tt
  arXiv:1104.0243}}] [\href{http://inspirebeta.net/record/894669}{Inspire}].

\bibitem{Buckley:2011ww}
M.~R. Buckley, D.~Hooper, J.~Kopp, and E.~Neil, {\it Light $z'$ {B}osons at the
  {T}evatron},  {\em Phys. Rev.} {\bf D 83} (2011) 115013,
  [\href{http://xxx.lanl.gov/abs/1103.6035}{{\tt arXiv:1103.6035}}] [\href{http://inspirebeta.net/record/894635}{Inspire}].

\bibitem{Cheung:2011un}
K.~Cheung and J.~Song, {\it Baryonic ${Z}^{\prime}$ {E}xplanation for the {CDF}
  $wjj$ {E}xcess},  {\em Phys. Rev. Lett.} {\bf 106} (2011) 211803,
  [\href{http://xxx.lanl.gov/abs/1104.1375}{{\tt arXiv:1104.1375}}] [\href{http://inspirebeta.net/record/895188}{Inspire}].

\bibitem{Wang:2011wh}
X.-P. Wang, Y.-K. Wang, B.~Xiao, J.~Xu, and S.~hua Zhu, {\it ${O}(100 {G}e{V})$
  {D}eci-weak ${W}^\prime/{Z}^\prime$ at {T}evatron and {LHC}},  {\em Phys.
  Rev.} {\bf D 83} (2011) 117701,
  [\href{http://xxx.lanl.gov/abs/1104.1161}{{\tt arXiv:1104.1161}}] [\href{http://inspirebeta.net/record/895065}{Inspire}].

\bibitem{Anchordoqui:2011wg}
L.~A. Anchordoqui, H.~Goldberg, X.~Huang, D.~Lust, and T.~R. Taylor, {\it
  Stringy origin of {T}evatron $wjj$ anomaly},  {\em Phys. Lett.} {\bf B 701}
  (2011) 224--228, [\href{http://xxx.lanl.gov/abs/1104.2302}{{\tt
  arXiv:1104.2302}}] [\href{http://inspirebeta.net/record/895781}{Inspire}].

\bibitem{Gunion:2011bx}
J.~F. Gunion, {\it A two-{H}iggs-doublet interpretation of a small {T}evatron
  ${W}jj$ excess},  \href{http://xxx.lanl.gov/abs/1106.3308}{{\tt
  arXiv:1106.3308}} [\href{http://inspirebeta.net/record/913819}{Inspire}].

\bibitem{Harnik:2011mv}
R.~Harnik, G.~D. Kribs, and A.~Martin, {\it {Q}uirks at the {T}evatron and
  {B}eyond},  {\em Phys. Rev.} {\bf D 84} (2011) 035029,
  [\href{http://xxx.lanl.gov/abs/1106.2569}{{\tt arXiv:1106.2569}}] [\href{http://inspirebeta.net/record/913637}{Inspire}].

\bibitem{Giudice:2011ak}
G.~F. Giudice, B.~Gripaios, and R.~Sundrum, {\it Flavourful {P}roduction at
  {H}adron {C}olliders},  {\em JHEP} {\bf 08} (2011) 055,
  [\href{http://xxx.lanl.gov/abs/1105.3161}{{\tt arXiv:1105.3161}}] [\href{http://inspirebeta.net/record/900111}{Inspire}].

\bibitem{DAmbrosio:2002kc}
G.~D'Ambrosio, G.~F. Giudice, G.~Isidori, and A.~Strumia, {\it {Minimal Flavour
  Violation: an effective field theory approach}},  {\em Nucl. Phys.} {\bf B
  645} (2002) 155--187, [\href{http://xxx.lanl.gov/abs/hep-ph/0207036}{{\tt
  hep-ph/0207036}}] [\href{http://inspirehep.net/record/589708}{Inspire}].

\bibitem{Vecchi:2011un}
L.~Vecchi, {\it Color $\&$ {W}eak triplet scalars, the dimuon asymmetry in
  ${B}_{s}$ decay, the top {FB} asymmetry, and the {CDF} dijet excess},  {\em
  JHEP} {\bf 10} (2011) 003, [\href{http://xxx.lanl.gov/abs/1107.2933}{{\tt
  arXiv:1107.2933}}] [\href{http://inspirebeta.net/record/918740}{Inspire}].

\bibitem{Nelson:2011up}
A.~E. Nelson, T.~Okui, and T.~S. Roy, {\it A unified, flavor symmetric
  explanation for the $t\bar{t}$ asymmetry and $wjj$ excess at {CDF}},
  \href{http://xxx.lanl.gov/abs/1104.2030}{{\tt arXiv:1104.2030}} [\href{http://inspirebeta.net/record/895647}{Inspire}].

\bibitem{CDFCollaboration:2009hy}
{\bf CDF} Collaboration, T.~Aaltonen {\em et.~al.}, {\it Search for new
  particles decaying into dijets in proton-antiproton collisions at
  $\sqrt{s}=1.96$ {T}e{V}},  {\em Phys. Rev.} {\bf D 79} (2009) 112002,
  [\href{http://xxx.lanl.gov/abs/0812.4036}{{\tt arXiv:0812.4036}}] [\href{http://inspirebeta.net/record/805902}{Inspire}].

\bibitem{afterlep2}
R.~Barbieri, A.~Pomarol, R.~Rattazzi, and A.~Strumia, {\it {Electroweak
  symmetry breaking after LEP-1 and LEP-2}},  {\em Nucl.Phys.} {\bf B703}
  (2004) 127--146, [\href{http://xxx.lanl.gov/abs/hep-ph/0405040}{{\tt
  hep-ph/0405040}}] [\href{http://inspirehep.net/record/649700}{Inspire}].

\bibitem{Alwall:2011uj}
J.~Alwall, M.~Herquet, F.~Maltoni, O.~Mattelaer, and T.~Stelzer, {\it
  Mad{G}raph 5: {G}oing {B}eyond},  {\em JHEP} {\bf 06} (2011) 128,
  [\href{http://xxx.lanl.gov/abs/1106.0522/hep-ph}{{\tt 1106.0522/hep-ph}}] [\href{http://inspirebeta.net/record/912611}{Inspire}].

\bibitem{UTfitCollaboration:2007p3234}
{\bf UTfit} Collaboration, M.~Bona {\em et.~al.}, {\it Model-independent
  constraints on ${\Delta} {F}=2$ operators and the scale of {N}ew {P}hysics},
  {\em JHEP} {\bf 03} (2008) 049,
  [\href{http://xxx.lanl.gov/abs/0707.0636}{{\tt arXiv:0707.0636}}] [\href{http://inspirebeta.net/record/755026}{Inspire}].

\bibitem{Dreiner:2008tw}
H.~K. Dreiner, H.~E. Haber, and S.~P. Martin, {\it Two-component spinor
  techniques and {F}eynman rules for quantum field theory and supersymmetry},
  {\em Phys. Rept.} {\bf 494} (2010) 1--196,
  [\href{http://xxx.lanl.gov/abs/0812.1594}{{\tt arXiv:0812.1594}}] [\href{http://inspirebeta.net/record/804666}{Inspire}].

\bibitem{Lunghi:2001af}
E.~Lunghi and D.~Wyler, {\it Complex flavor couplings in supersymmetry and
  unexpected {CP} violation in the decay ${B} \to \phi {K}$},  {\em Phys.
  Lett.} {\bf B 521} (2001) 320--328,
  [\href{http://xxx.lanl.gov/abs/hep-ph/0109149}{{\tt hep-ph/0109149}}] [\href{http://inspirebeta.net/record/562899}{Inspire}].

\bibitem{Hurth:2008p3281}
T.~Hurth, G.~Isidori, J.~F. Kamenik, and F.~Mescia, {\it Constraints on {N}ew
  {P}hysics in {MFV} models: a model-independent analysis of ${\Delta} f=1$
  processes},  {\em Nucl. Phys.} {\bf B 808} (2009) 326--346,
  [\href{http://xxx.lanl.gov/abs/0807.5039}{{\tt arXiv:0807.5039}}] [\href{http://inspirebeta.net/record/791900}{Inspire}].

\bibitem{Low:2009gl}
I.~Low, R.~Rattazzi, and A.~Vichi, {\it Theoretical {C}onstraints on the
  {H}iggs {E}ffective {C}ouplings},  {\em JHEP} {\bf 04} (2010) 126,
  [\href{http://xxx.lanl.gov/abs/0907.5413}{{\tt arXiv:0907.5413}}] [\href{http://inspirebeta.net/record/827541}{Inspire}], and the references
  cited therein.

\bibitem{Aad:2011yh}
{\bf ATLAS} Collaboration, G.~Aad {\em et.~al.}, {\it Search for {M}assive
  {C}olored {S}calars in {F}our-{J}et {F}inal {S}tates in $\sqrt{s}=7$ {T}e{V}
  proton-proton collisions with the {ATLAS} {D}etector},
  \href{http://xxx.lanl.gov/abs/1110.2693}{{\tt arXiv:1110.2693}} [\href{http://inspirehep.net/record/939560}{Inspire}].

\bibitem{DelNobile:1199750}
E.~D. Nobile, R.~Franceschini, D.~Pappadopulo, and A.~Strumia, {\it {M}inimal
  {M}atter at the {L}arge {H}adron {C}ollider},  {\em Nucl. Phys.} {\bf B 826}
  (2009) 217--234, [\href{http://xxx.lanl.gov/abs/0908.1567}{{\tt
  arXiv:0908.1567}}] [\href{http://inspirebeta.net/record/828294}{Inspire}].

\bibitem{annovi_private}
A.~Annovi, P.~Catastini, V.~Cavaliere, and L.~Ristori, {\it Private
  communication}.

\bibitem{Campbell:387805}
J.~M. Campbell and R.~K. Ellis, {\it An update on vector boson pair production
  at hadron colliders},  {\em Phys. Rev.} {\bf D60} (1999) 113006,
  [\href{http://xxx.lanl.gov/abs/hep-ph/9905386}{{\tt hep-ph/9905386}}] [\href{http://inspirebeta.net/record/500112}{Inspire}].

\end{thebibliography}\endgroup
\bibliographystyle{JHEP}


\end{document}